\documentclass[10pt,twocolumn,english,a4paper, prl]{revtex4}
\usepackage{mathptmx}
\usepackage[T1]{fontenc}
\usepackage[latin9]{inputenc}
\usepackage[a4paper]{geometry}
\usepackage{amsmath}
\usepackage{graphicx}
\usepackage{amssymb}

\makeatletter

\providecommand{\tabularnewline}{\\}

\@ifundefined{textcolor}{}
{%
 \definecolor{BLACK}{gray}{0}
 \definecolor{WHITE}{gray}{1}
 \definecolor{RED}{rgb}{1,0,0}
 \definecolor{GREEN}{rgb}{0,1,0}
 \definecolor{BLUE}{rgb}{0,0,1}
 \definecolor{CYAN}{cmyk}{1,0,0,0}
 \definecolor{MAGENTA}{cmyk}{0,1,0,0}
 \definecolor{YELLOW}{cmyk}{0,0,1,0}
 }

\usepackage{braket}

\makeatother

\usepackage{babel}

\begin{document}

\title{Symmetry-induced gap opening in graphene superlattices }

\author{Rocco Martinazzo, Simone Casolo and Gian Franco Tantardini }

\affiliation{Department of Physical Chemistry and Electrochemistry, University
of Milan, V. Golgi 19, 20133 Milan, Italy}

\affiliation{CIMAINA, Interdisciplinary Center of Nanostructured Materials and
Interfaces, University of Milan. E-mail: rocco.martinazzo@unimi.it}
\begin{abstract}
We study $n$x$n$ honeycomb superlattices of defects in graphene.
The considered defects are missing $p_{z}$ orbitals and can be realized
by either introducing C atom vacancies or chemically binding simple
atomic species at the given sites. Using symmetry arguments we show
how it is possible to open a gap when $n=3m+1,3m+2$ ($m$ integer),
and estimate its value to have an approximate square-root dependence
on the defect concentration $x=1/n^{2}$. Tight-binding calculations
confirm these findings and show that the induced-gaps can be quite
large, e.g. $\sim100$ meV for $x\sim10^{-3}$. Gradient-corrected
density functional theory calculations on a number of superlattices
made by H atoms adsorbed on graphene are in good agreement with tight-binding
results, thereby suggesting that the proposed structures may be used
in practice to open a gap in graphene.
\end{abstract}
\maketitle
The recent fabrication of graphene\citep{novoselov04}, a one atom-thick
layer of carbon atoms arranged in a honeycomb lattice, has triggered
a wealth of studies in both fundamental and applied science. Graphene
is a a zero-gap semiconductor with a linear dispersion at the Fermi
level in which low-energy excitations mimic the behaviour of relativistic
massless fermions\citep{novoselov05,zhang05} (see \citep{geim07,castroneto09}
for reviews). This gives rise to a number of interesting phenomena
such as an anomalous quantum Hall effect\citep{novoselov05,zhang05}
and quasirelativistic Klein tunneling\citep{katsnelson06}. Its unconventional
transport properties with ballistic transport on submicrometer scale\citep{schedin07}
and with carrier mobilities up to $2\,10^{5}$ $cm^{2}V^{-1}s^{-1}$
\citep{bolotin08} offer the possibility of high-performance interconnects
in an hypothetical carbon-based nanoelectronics. However, since conductivity
cannot be turned completely off, pristine graphene cannot be used
as a transistor in logic applications, where high on/off ratios are
required\citep{avouris07,novolesov07}. Field-switching capabilities
depend on the presence (and size) of a gap in the electronic structure.
Electron confinement can be used to open a gap inversely proportional
to the confinement length when rolling up graphene into single-walled
nanotubes or cutting its edges to form nanoribbons. Both these possibilities
have been exploited, and promising carbon nanotube/nanoribbons field-effect
transistors realized\citep{avouris07}. 

In this Letter we explore a different possibility made available by
recent progresses in patterning graphene with lithographic techniques\citep{meyer08,fischbein08}.
Specifically, we study $n$x$n$ honeycomb superlattices of defects
on graphene (defined to be periodic structures of defects arranged
to form a honeycomb lattice commensurate with the substrate) and use
symmetry arguments to design semiconducting structures. We show that
a gap can be opened by preserving graphene symmetry. We estimate that
an inverse proportionality to the (super)lattice constant $n$ approximately
holds for the gap size in these structures, and use tight-binding
and \emph{first}-\emph{principles} calculations to validate these
predictions. The computed gaps are sizable and compare favourably
with those found in nanoribbons at the same length scale\citep{louie06}.
In the following we first present our symmetry arguments and derive
general rules to open a gap in the considered superlattices. We then
estimate its value and present the results of numerical calculations. 

Graphene's unconventional electronic properties are strictly related
 to its $D_{6h}$ point symmetry. The $k-$group at the $K$ ($K'$)
high-symmetry points ($D_{3h}$) allows for doubly degenerate irreducible
representations, and Bloch functions built with $p_{z}$ orbitals
of the $A$ and $B$ sublattices span one of its two-dimensional irreducible
representation (irrep), namely $E^{\prime\prime}$. This is enough
for the $\pi-\pi^{*}$ degeneracy at the $K$($K'$) point and for
the unusual linear dispersion relation, irrespective of the level
of approximation\citep{Weiss58}. These properties are captured by
the simple tight-binding (TB) model Hamiltonian\begin{equation}
H=-t\sum_{<i,j>}a_{i}^{\dagger}b_{j}+H.c.-t'\sum_{\ll i,j\gg}a_{i}^{\dagger}a_{j}-t'\sum_{\ll i,j\gg}b_{i}^{\dagger}b_{j}\label{eq:TB hamiltonian}\end{equation}
where $a_{i}^{\dagger}$($b_{i}^{\dagger}$) is the creation operator
for an electron on site $i$ of the A (B) sublattice, the first sums
run over nearest neighboring sites in the honeycomb lattice and the
second sums run over sites which are nearest neighbors in the triangular
sublattices. The hopping $t$ has been estimated to be $\sim2.7$
eV whereas $t'\ll t$ has different values depending on the parametrization.
When $t'=0$ the Hamiltonian describes a bipartite system, and we
assume that this approximately holds for graphene. The consequences
of relaxing this approximation will be addressed numerically at the
end of this Letter. 

Bipartitism has a large impact on the electronic structure \emph{via}
the induced electron-hole symmetry. For instance, it has long been
known that in bipartite systems, at half filling, sublattice imbalances
due to vacancies strongly affect the energy spectrum at the Fermi
level through the introduction of midgap states\citep{Inui94}. In
graphene, such states have a semidelocalized nature with a $1/r$
dependence on the distance from the defect\citep{Pereira2006}. By
introducing an equal number of defects on each sublattice one restores
balance, eliminates midgap states and a gap possibly opens. In general,
however, there is no guarantee that the gap opens at $K$ and does
not close somewhere else in the Brillouin zone (BZ). Therefore, we
focus here on $n$x$n$ honeycomb patterns of defects only, in such
a way to constrain (by symmetry) the changes in the band structure
and possibly reduce accidental degeneracies. In these structures the
high-symmetry points where degeneracy is expected are the $\Gamma$
and $K$ points, that is where the $k$-groups ($D_{6h}$ and $D_{3h}$,
respectively) allow for doubly degenerate irreps. In the following
we use $E$ ($A$) for a generic two- (one-) dimensional irreducible
representation, and denote as $K_{n}$ the $K$ point of the $n$x$n$
superlattice BZ (SBZ). For a strictly bipartite system at half-filling,
degeneracy at the Fermi level occurs when the number of $E$ irreps
is odd; when this number is even degeneracy, if occurs, can be considered
as accidental. Therefore, introducing symmetric defects in such a
way to have an even number of $E$ irreps \emph{both} at the $\Gamma$
and at the $K$ point a gap generally opens. 

\begin{figure}
\begin{centering}
\includegraphics[clip,width=0.6\columnwidth]{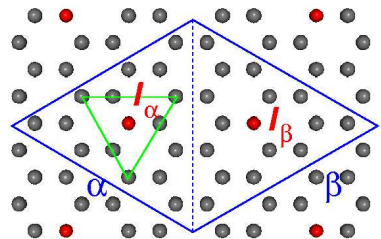}
\par\end{centering}

\caption{\label{fig:cell}The simplest symmetrically defective superlattice
for $n=4$. Also indicated the two halves of the cell ($\alpha$,$\beta$),
their centers $I_{\alpha},I_{\beta}$ (red balls) and a triangle $\Delta_{i,\alpha}$.
A defect is a missing $p_{z}$ orbital at $I_{\alpha},I_{\beta}$. }

\end{figure}

To show that this is indeed possible, we consider a generic $n$x$n$
supercell and count the number of $A$ and $E$ irreps generated by
the carbon atoms in cell (the so-called atomic representations). To
this end, it is sufficient to consider that half of the cell which
has $D_{3h}$ symmetry with respect to its center $I$ (see Fig.\ref{fig:cell}),
the remaining half behaving similarly. These two half-cells ($\alpha$
and $\beta$ in the following) play the role of A and B type of sites
in the honeycomb superlattice. In each of them, the set of C atoms
may be grouped in classes of equilateral triangles $\Delta_{i,\alpha}$
($\Delta_{i,\beta}$), plus a possible atom at $I_{\alpha}$($I_{\beta}$)
as it happens when $n=3m+1$ and $n=3m+2$ ($m$ integer). Each triangle
spans an $A+E$ irrep of the $D_{3h}$ group (centered at $I$) which
behave as $s$ and ($p_{x}$,$p_{y}$) orbitals centered at $I$;
the atom at $I$, when present, spans of course an $A$ irrep. Then,
by considering Bloch functions built with these $s$- and $(p_{x},p_{y}$)-
like orbitals it is possible to count the number of $A$ and $E$
irreps for each case. At $\Gamma$ the Bloch functions built with
$s$-like orbitals centered on $I_{\alpha}$ and $I_{\beta}$ span
two $A$ representations, whereas $p_{x},p_{y}$-like functions span
two $E$ irreps; at $K_{n}$ the first generate an $E$ irrep whereas
the latter span $2A+E$. These are also the irreps generated by the
$p_{z}$ orbitals of the C atoms as long as we discriminate between
$A$ and $E$ type only. The overall result is given in Table \ref{tab:Irreps}
where the symbol $\bar{n}_{3}=\bar{0}_{3},\bar{1}_{3},\bar{2}_{3}$
identifies the three (congruence) classes modulo $3$, i.e. the sequences
$n=3m,3m+1$ and $3m+2$, respectively. %
\begin{table}
\begin{centering}
\begin{tabular}{|c|c|c|}
\hline 
$\Gamma$ & A & E\tabularnewline
\hline 
$\bar{0}_{3}$ & $2n^{2}$ & $2n^{2}$\tabularnewline
\hline 
$\bar{1}_{3}$ & $2(3n^{2}+2n+1)$ & $2(3n^{2}+2n)$\tabularnewline
\hline 
$\bar{2}_{3}$ & $2(3n^{2}+4n+2)$ & $2(3n^{2}+4n+1)$\tabularnewline
\hline
\hline 
$K_{n}$ & A & E\tabularnewline
\hline 
$\bar{0}_{3}$ & $2n^{2}$ & $2n^{2}$\tabularnewline
\hline 
$\bar{1}_{3}$ & $2n(3n+2)$ & $2n(3n+2)+1$\tabularnewline
\hline 
$\bar{2}_{3}$ & $2(3n^{2}+4n+1)$ & $2(3n^{2}+4n+1)+1$\tabularnewline
\hline
\end{tabular}
\par\end{centering}

\begin{centering}
\caption{\label{tab:Irreps}Number of irreducible one- ($A$) and two- ($E$)
dimensional representations (per cell) generated by the full atomic
basis on a $n$x$n$ honeycomb superlattice at the $\Gamma$ and K
points of the corresponding BZ. }

\par\end{centering}

\end{table}

It follows from Table \ref{tab:Irreps} that with the full atomic
set (i.e. considering pure graphene) degeneracy occurs at the $K_{n}$
points when either $n\in\bar{1}_{3}$ or $n\in\bar{2}_{3}$. This
is consistent with the folding $K(K')\rightarrow K_{n}(K'_{n})$ and
$K(K')\rightarrow K'_{n}(K_{n})$, respectively. In the case $n\in\bar{0}_{3}$
both $K$ and $K'$ folds to $\Gamma$ and therefore a 4-fold degeneracy
occurs; this can be considered accidental in this context as it cannot
be predicted by the number of $E$ irreps only. More interestingly,
two important results concerning the introduction of $p_{z}-$vacancies
are easily proved. 

I. \emph{By removing a $\Delta_{i,\alpha}$, $\Delta_{i,\beta}$ pair
only is }not\emph{ generally possible to open a gap.} Here, $2(A+E)$
irreps are removed both at $\Gamma$ and at $K_{n}$, and no modification
occurs on the parity of the $E$ sets. Exceptions to this rule are,
of course, those cases where degeneracy is accidental (as pure graphene
in the $\bar{0}_{3}$ case which does show a gap after removal of
one such pairs, see below). 

II.\emph{ When }$n\in\bar{1}_{3}$ or $n\in\bar{2}_{3}$\emph{ removal
of the atoms at $I_{\alpha}$ and $I_{\beta}$ does open a gap.} In
these cases, the atomic basis spans $2A$ at $\Gamma$ and $E$ at
$K_{n}$, thereby turning the number of $E$ irreps to be even at
\emph{both} special points. Also in this case, exceptions of residual
accidental degeneracy are possible. 

The second result provides a very simple way for opening a gap in
two thirds of the cases, that is by introducing a $p_{z}$ vacancy
at $I_{\alpha}$ and $I_{\beta}$. In practice, this can be realized
by either removing substrate atoms or using them to covalently bind
simple adsorbates such as H atoms\citep{casolo09}. In the latter
case, indeed, the C atoms involved in the chemisorption process turn
their hybridization to $sp^{3}$ and effectively get out of the $\pi-\pi^{*}$
band system. In this Letter we focus on the simplest defective superlattices,
i.e. those with vacancies at $I_{\alpha}$ and $I_{\beta}$ \emph{only},
and call them $(n,0)$-honeycombs. In general, the pair of integers
$(n,p)$ ($p=0,..\mathsf{int}(n^{2}/3)$) can be used to identify
a $nxn$ honeycomb superlattice with $2p$ equilateral triangles symmetrically
removed from the unit supercell, in addition to the atoms at their
centers $I_{\alpha}$ and $I_{\beta}$ when $n\in\bar{1}_{3}\bigcup\bar{2}_{3}$
. Fig.\ref{fig:cell} shows an example, the $(4,0)$-honeycomb. 

\begin{figure}
\begin{centering}
\includegraphics[clip,width=1\columnwidth]{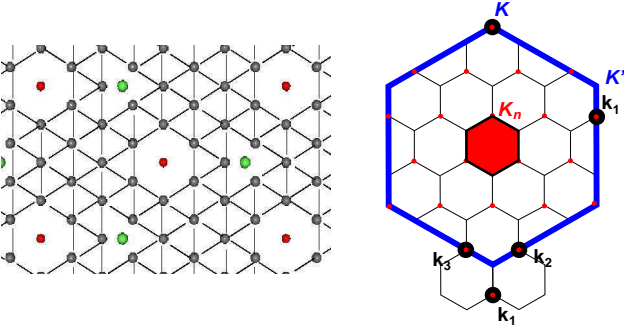} 
\par\end{centering}

\caption{\label{fig:renormalized latticel}Left panel: the renormalized triangular
lattice corresponding to the $(4,0)$-honeycomb (grey balls). Red
and green balls mark the position of the A and B defects. Right panel:
Brillouin zones for grahene (blue borders) and for the structure shown
on the left panel (red). Also indicated the four lowest-energy $\mathbf{k}$
points ($K$ and $\mathbf{k}_{i}$ $i=1-3$) for the calculation in
the text (black circles). }

\end{figure}
It is possible to estimate the size of the induced gap at the $K_{n}$
point of these $(n,0)$-honeycombs. To this end, we perform a lattice
renormalization\citep{naumis07} by making use of the bipartite nature
of the Hamiltonian $H=H_{AB}+H_{BA}$. This simplifies the problem
by halving the state space of interest. Indeed, since $H$ only allows
transitions from the A to the B subspaces ($H_{BA}$) and \emph{vice}
\emph{versa} ($H_{BA}$) it is sufficient to consider the problem
in the A space only with Hamiltonian%
\footnote{If $H$ is the second quantization version of the Hamiltonian a projection
onto the single particle space is implied.%
} $\tilde{H}_{AA}=H_{AB}H_{BA}$. For any non-zero eigenvalue $\tilde{\epsilon}_{i}$
and eigenvector $\ket{\psi_{A,i}}$ of this Hamiltonian there exist
two solutions of the original problem with eigenvalues $\epsilon_{i}=\pm\sqrt{\tilde{\epsilon}_{i}}$
and eigenvectors $\ket{\psi_{A,i}}\pm\ket{\psi_{B,i}}$, where $\ket{\psi_{B,i}}$
is defined to be $\ket{\psi_{B,i}}=\tilde{\epsilon}_{i}^{-1/2}H_{BA}\ket{\psi_{A,i}}$;
if $\tilde{\epsilon}_{i}=0$, $\ket{\psi_{A,i}}$ is already a $H$
eigenvector. The converse is also true, namely from any eigenvector
$\ket{\psi_{i}}$ the two projections $\ket{\psi_{A,i}}$ and $\ket{\psi_{B,i}}$
onto the A and B supspaces satisfy $H_{BA}\ket{\psi_{A,i}}=\epsilon_{i}\ket{\psi_{B,i}}$
and $\tilde{H}_{AA}\ket{\psi_{A,i}}=\epsilon_{i}^{2}\ket{\psi_{A,i}}$;
that is, in studying $\tilde{H}_{AA}$ one only misses possible zero
eigenstates in the B subspace%
\footnote{Their presence can easily be detected by defining A to be the majority
species and comparing the number of zero A eigenstates with the sublattices
inbalance.%
}. In graphene the renormalized Hamiltonian $\tilde{H}_{AA}$ describes
a triangular lattice with on-site energy $t^{2}Z$ (where $t$ is
the hopping term of eq.\ref{eq:TB hamiltonian} and $Z=3$ is the
coordination number of A atoms in the original honeycomb lattice)
and hopping $t^{2}$ between neighbors in the triangular lattice.
Defects are of two kinds: while A vacancies translate simply into
A vacancies in the renormalized lattice, B vacancies modify both the
coordination number and the hopping between A sites. The renormalized
$(n,0)$-honeycomb for $n=4$ is shown in the left panel of Fig.\ref{fig:renormalized latticel},
whereas the right panel of the same figure displays its SBZ along
with the graphene BZ. The state space at $K_{n}$ is given by its
$n^{2}$ replicas within BZ and comprises $K$ or $K'$ depending
on whether $n\in\bar{1}_{3}$ or $n\in\bar{2}_{3}$. Quasi-degenerate
perturbation theory is necessary to estimate the ground-state energy
$\tilde{\epsilon}_{0}(K_{n})$ (and hence the gap $\epsilon_{gap}=2\sqrt{\tilde{\epsilon}_{0}}$),
but it becomes intractable at large $n$ (i.e. at very small defect
concentration $x=1/n^{2}$) because a huge number of $K_{n}$ replicas
gets close to the $K$ and $K'$ points. Therefore, we consider $x$
sufficiently small that the defects are isolated from each other,
but large enough that a few state calculation is reliable. The smallest
set of $K_{n}$ replicas contains $K$ and the $\mathbf{k}_{i}$ vectors
($i=1-3$) shown in Fig.\ref{fig:renormalized latticel} (right panel)
for the case $n=4$, and corresponds to the set of Bloch functions
$\ket{\psi_{0}}=\ket{\psi_{K}}$ and $\ket{\psi_{i}}=\ket{\psi_{\mathbf{k_{i}}}}$
built with $p_{z}$ orbitals at A sites; the case $n\in\bar{2}_{3}$
is analogous except that $K$ is replaced with $K'$. $\ket{\psi_{0}}$
spans the $A_{2}^{\prime\prime}$ irrep of the $k$-group at $K_{n}$
($D_{3h}$), whereas $\{\ket{\psi_{i}}\}$ span $A_{2}^{\prime\prime}+E^{\prime\prime}$.
Thus it is possible to set up a two dimensional problem in the $A_{2}^{\prime\prime}$
subspace. The corresponding Hamiltonian matrix can be obtained from
the $\tilde{H}_{AA}$ matrix elements between (graphene) Bloch states,
Eq.\ref{eq:kk' matrix elements}

\begin{align}
\braket{\psi_{\mathbf{k'}}\mathbf{|}\tilde{H}_{AA}|\psi_{\mathbf{k}}} & =t^{2}\delta_{\mathbf{k},\mathbf{k}'}(3+F(\mathbf{k}))\nonumber \\
 & -xt^{2}\delta_{\mathbf{k},\mathbf{k}'+\mathbf{g}}\left\{ e^{i\mathbf{g}\boldsymbol{\delta}_{A}}(3+F(\mathbf{k}')+F(\mathbf{k}))\right.\label{eq:kk' matrix elements}\\
 & \left.+e^{i\mathbf{g}\boldsymbol{\delta}_{B}}f(\mathbf{k}')^{*}f(\mathbf{k})\right\} \nonumber \end{align}
\begin{figure}
\begin{centering}
\includegraphics[clip,width=0.8\columnwidth]{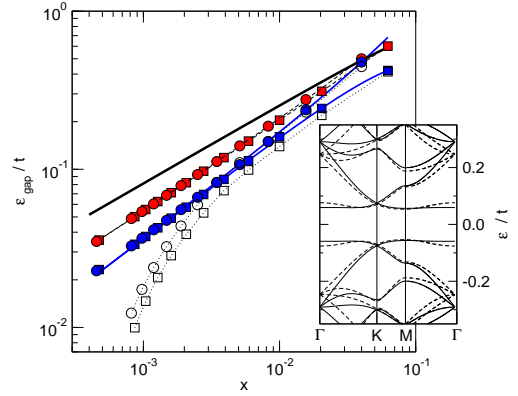}
\par\end{centering}

\caption{\label{fig:TB gaps}Energy gaps (in units of $t$) for the $(n,0)$-honeycombs
as functions of $x=1/n^{2}$. Bold black line is the gap at $K_{n}$
as estimated using Eq. \ref{eq:2x2matrix}. Filled symbols are for
results of tight-binding calculations with $t'=0$ and bold blue lines
are best-fit curves (see text): squares for $n\in\bar{1}_{3}$, circles
for $n\in\bar{2}_{3}$; red symbols for the gap at $K_{n}$, blue
for the full gap. Open symbols are full gap results for $t'=0.1\, t$.
The inset shows the low-energy band structures of the $(13,0)$ (solid
lines) and $(14,0)$ (dashed lines) honeycombs as obtained from tight-binding
calculations with $t'=0$. }

\end{figure}
Here $F(\mathbf{k})=\sum_{i=1}^{6}e^{-i\mathbf{k}\boldsymbol{\delta}_{i}'}$,
$f(\mathbf{k})=\sum_{i=1}^{3}e^{-i\mathbf{k}\boldsymbol{\delta}_{i}}$
( where $\boldsymbol{\delta}_{i}'$ and $\boldsymbol{\delta}_{i}$
are the vectors joining AA and AB nearest-neighbors, respectively
), $\mathbf{g}$ is a reciprocal superlattice vector and \textbf{$\boldsymbol{\delta}_{A},\boldsymbol{\delta}_{B}$
}are the position vectors of the defects in the unit supercell. In
deriving Eq. \ref{eq:kk' matrix elements} periodicity of the superlattice
has been used and the defects have been considered as isolated ($n>2$).
With the help of Eq. \ref{eq:kk' matrix elements} and of the symmetry
properties of $F(\mathbf{k})$ and $f(\mathbf{k})$,  the Hamiltonian
matrix in the above $A_{2}^{\prime\prime}$ space reads as \begin{equation}
t^{2}\left[\begin{array}{ccc}
3x &  & -x\sqrt{3}F_{x}\\
-x\sqrt{3}F_{x} &  & 3+F_{x}-9x(2+F_{x})\end{array}\right]\label{eq:2x2matrix}\end{equation}
where $F_{x}=F(\mathbf{k}_{1})\sim-3+3(2\pi/3)^{2}x-\sqrt{3}(2\pi/3)^{3}x^{3/2}$.
The lowest eigenvalue $\tilde{\epsilon}_{0}$ ($\sim3xt^{2}\,(0.561-0.961\sqrt{x})$
) allows us to estimate the energy gap $\epsilon_{gap}$ at the $K_{n}$
point. This is plotted in Fig.\ref{fig:TB gaps} as a function of
$x$ along with the results of tight-binding calculations. The importance
of including a larger number of $K_{n}$ replicas close to ($K$,
$K'$) is evident at small $x$ where the energy gap decreases slightly
faster than $\sqrt{x}$; the best-fit to the numerical results gives
an exponent $\sim0.66$. Tight-binding calculations reveal that the
minimum gaps occurs either at $M$ or $\Gamma$ (see inset of Fig.\ref{fig:TB gaps}),
depending on $x$ and on the sequence considered, but behave similarly
to the gaps at $K_{n}$. Differences between the two sequences appear
at large $x$ and reflects the different shape of the low-energy bands
(not shown). A best-fit of $\epsilon_{gap}=ax^{\alpha}(1+bx^{\beta})$
to the numerical results, also reported in Fig.\ref{fig:TB gaps},
gives $(a,b,\alpha,\beta)=(3.34,-4.99,0.65,1.14)$ for $n\in\bar{1}_{3}$
and $(3.37,+2.9,0.66,0.88)$ for $n\in\bar{2}_{3}$. We have also
investigated the effect of breaking the electron-hole symmetry by
performing tight-binding calculations with $t'\neq0$. Introduction
of the next-to-nearest neighbors interaction only affects the results
at small $x$, where the valence and conduction bands start to overlap
at some point $x_{c}$ because of the asymmetry introduced in the
energy spectrum. The position of this critical value $x_{c}$ shifts
to larger $x$ when increasing $t'$ but remains small for realistic
values (for $t'=0.1\, t$ used in Fig.\ref{fig:TB gaps}  $x_{c}<10^{-3}$).
Notice that the effect of next-to-nearest neighbor interactions is
different for the two sequences considered. %
\begin{figure}
\begin{centering}
\includegraphics[clip,width=0.8\columnwidth]{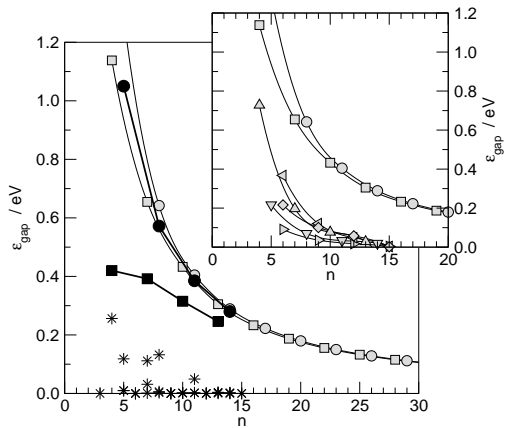}
\par\end{centering}

\caption{\label{fig:DFT gaps}DFT results for the energy gap in $(n,0)$-honeycombs
made by adsorbed H atoms as functions of $n$ (filled symbols). For
comparison, also reported are the tight-binding results of Fig. \ref{fig:TB gaps}
(open symbols) with $t=2.7$ eV and $t'=0$. Squares for $n\in\bar{1}_{3}$,
circles for $n\in\bar{2}_{3}$. Stars are energy gaps in a number
of superlattices with an asymmetric AB pair of defects per cell (\emph{ortho-}
and \emph{para-} dimers \emph{not} included). The inset shows TB results
(normalized to the number of pairs of defects) for a number of $(n,p)$
honeycombs: triangles for $p=1$, diamonds for $p=2$. (Lines are
guide to eyes). }

\end{figure}

To further investigate this point and, more importantly, to address
the role of electron correlation we performed gradient-corrected density-functional-theory
(DFT) calculations on a number of $(n,0)$-honeycombs of adsorbed
H atoms. The details of the calculations are analogous to those reported
in a previous work\citep{casolo09}, except for the fact that a finer
mesh of $\mathbf{k}$ points has been used to correctly compute the
energy gaps. The results are reported in Fig.\ref{fig:DFT gaps} up
to $n=14$ ($x\sim0.005$); for larger values of $n$ DFT calculations
become prohibitive. As can be seen from Fig. \ref{fig:DFT gaps} DFT
results show a reduced gap size with respect to TB ones in all the
cases considered but the effect is much more pronounced in the $\bar{1}_{3}$
sequence than in the $\bar{2}_{3}$ one. In particular, DFT results
for $n=5,8,11,14$ closely parallel the TB ones ($t$ has been set
to its accepted value, $t=2.7$ eV), despite the fact that the first
refer to a realistic situation where defects are H atoms while in
TB calculations defects are modeled by simple $p_{z}$ vacancies.
Discrepancies in the $\bar{1}_{3}$ sequence at small $n$ needs further
investigation, though it is in line with the introduction of the next-to-nearest
neighbor hopping (Fig. \ref{fig:TB gaps}). Similar behaviour was
found for the gap in armchair graphene nanoribbons where it was explained
by the modified nearest-neighbor hoppings for the sites close to the
defect\citep{louie06}. In any case, both the size of the gap and
its dependence on $n$ are promising for future applications. 

Finally, to underline the role played by symmetry in designing the
defective structures, TB results for different superlattices are also
reported in Fig. \ref{fig:DFT gaps}. Notice that the $(n,0)$ honeycombs
considered in this Letter show the largest gaps with the minimum number
of defects per supercell (see the inset of Fig.\ref{fig:DFT gaps},
where the results for a number of $(n,p)$-honeycombs with $n\in\bar{0}_{3}$
are also reported). 

To summarize, we have studied graphene superlattices of defects where
a gap at the Fermi level opens because of symmetry. Tight-binding
and density-functional-theory calculations show that the gap is indeed
sizable, thereby suggesting that the proposed structures may play
a role in designing future carbon-based electronic devices. 

\bibliographystyle{apsrev}

\begin{thebibliography}{18}
\expandafter\ifx\csname natexlab\endcsname\relax\def\natexlab#1{#1}\fi
\expandafter\ifx\csname bibnamefont\endcsname\relax
  \def\bibnamefont#1{#1}\fi
\expandafter\ifx\csname bibfnamefont\endcsname\relax
  \def\bibfnamefont#1{#1}\fi
\expandafter\ifx\csname citenamefont\endcsname\relax
  \def\citenamefont#1{#1}\fi
\expandafter\ifx\csname url\endcsname\relax
  \def\url#1{\texttt{#1}}\fi
\expandafter\ifx\csname urlprefix\endcsname\relax\def\urlprefix{URL }\fi
\providecommand{\bibinfo}[2]{#2}
\providecommand{\eprint}[2][]{\url{#2}}

\bibitem[{\citenamefont{Novoselov et~al.}(2004)\citenamefont{Novoselov, Geim,
  Morozov, Jiang, Zhang, Dubonos, Gregorieva, and Firsov}}]{novoselov04}
\bibinfo{author}{\bibfnamefont{K.~S.} \bibnamefont{Novoselov}},
  \bibinfo{author}{\bibfnamefont{A.~K.} \bibnamefont{Geim}},
  \bibinfo{author}{\bibfnamefont{S.~V.} \bibnamefont{Morozov}},
  \bibinfo{author}{\bibfnamefont{D.}~\bibnamefont{Jiang}},
  \bibinfo{author}{\bibfnamefont{Y.}~\bibnamefont{Zhang}},
  \bibinfo{author}{\bibfnamefont{S.~V.} \bibnamefont{Dubonos}},
  \bibinfo{author}{\bibfnamefont{I.~V.} \bibnamefont{Gregorieva}},
  \bibnamefont{and} \bibinfo{author}{\bibfnamefont{A.~A.}
  \bibnamefont{Firsov}}, \bibinfo{journal}{Science}
  \textbf{\bibinfo{volume}{306}}, \bibinfo{pages}{666} (\bibinfo{year}{2004}).

\bibitem[{\citenamefont{Novoselov et~al.}(2005)\citenamefont{Novoselov, Geim,
  Morozov, Jiang, Katsnelson, Gregorieva, Dubonos, and Firsov}}]{novoselov05}
\bibinfo{author}{\bibfnamefont{K.~S.} \bibnamefont{Novoselov}},
  \bibinfo{author}{\bibfnamefont{A.~K.} \bibnamefont{Geim}},
  \bibinfo{author}{\bibfnamefont{S.~V. S.~V.} \bibnamefont{Morozov}},
  \bibinfo{author}{\bibfnamefont{D.}~\bibnamefont{Jiang}},
  \bibinfo{author}{\bibfnamefont{M.~I.} \bibnamefont{Katsnelson}},
  \bibinfo{author}{\bibfnamefont{I.~V.} \bibnamefont{Gregorieva}},
  \bibinfo{author}{\bibfnamefont{S.~V.} \bibnamefont{Dubonos}},
  \bibnamefont{and} \bibinfo{author}{\bibfnamefont{A.~A.}
  \bibnamefont{Firsov}}, \bibinfo{journal}{Nature}
  \textbf{\bibinfo{volume}{438}}, \bibinfo{pages}{197} (\bibinfo{year}{2005}).

\bibitem[{\citenamefont{Zhang et~al.}(2005)\citenamefont{Zhang, Tan, Stormer,
  and Kim}}]{zhang05}
\bibinfo{author}{\bibfnamefont{Y.}~\bibnamefont{Zhang}},
  \bibinfo{author}{\bibfnamefont{Y.-W.} \bibnamefont{Tan}},
  \bibinfo{author}{\bibfnamefont{H.~L.} \bibnamefont{Stormer}},
  \bibnamefont{and} \bibinfo{author}{\bibfnamefont{P.}~\bibnamefont{Kim}},
  \bibinfo{journal}{Nature} \textbf{\bibinfo{volume}{438}},
  \bibinfo{pages}{201} (\bibinfo{year}{2005}).

\bibitem[{\citenamefont{Geim and Novoselov}(2007)}]{geim07}
\bibinfo{author}{\bibfnamefont{A.~K.} \bibnamefont{Geim}} \bibnamefont{and}
  \bibinfo{author}{\bibfnamefont{K.~S.} \bibnamefont{Novoselov}},
  \bibinfo{journal}{Nat. Mater.} \textbf{\bibinfo{volume}{6}},
  \bibinfo{pages}{183} (\bibinfo{year}{2007}).

\bibitem[{\citenamefont{{Castro Neto} et~al.}(2009)\citenamefont{{Castro Neto},
  Guinea, Peres, Novoselov, and Geim}}]{castroneto09}
\bibinfo{author}{\bibfnamefont{A.~H.} \bibnamefont{{Castro Neto}}},
  \bibinfo{author}{\bibfnamefont{F.}~\bibnamefont{Guinea}},
  \bibinfo{author}{\bibfnamefont{N.~M.~R.} \bibnamefont{Peres}},
  \bibinfo{author}{\bibfnamefont{K.~S.} \bibnamefont{Novoselov}},
  \bibnamefont{and} \bibinfo{author}{\bibfnamefont{A.~K.} \bibnamefont{Geim}},
  \bibinfo{journal}{Rev. Mod. Phys.} \textbf{\bibinfo{volume}{81}},
  \bibinfo{pages}{109} (\bibinfo{year}{2009}).

\bibitem[{\citenamefont{Katsnelson et~al.}(2006)\citenamefont{Katsnelson,
  Novoselov, and Geim}}]{katsnelson06}
\bibinfo{author}{\bibfnamefont{M.~I.} \bibnamefont{Katsnelson}},
  \bibinfo{author}{\bibfnamefont{K.~S.} \bibnamefont{Novoselov}},
  \bibnamefont{and} \bibinfo{author}{\bibfnamefont{A.~K.} \bibnamefont{Geim}},
  \bibinfo{journal}{Nature Phys.} \textbf{\bibinfo{volume}{2}},
  \bibinfo{pages}{620} (\bibinfo{year}{2006}).

\bibitem[{\citenamefont{Schedin et~al.}(2007)\citenamefont{Schedin, Geim,
  Morozov, Hill, Blake, Katsnelson, and Novoselov}}]{schedin07}
\bibinfo{author}{\bibfnamefont{F.}~\bibnamefont{Schedin}},
  \bibinfo{author}{\bibfnamefont{A.~K.} \bibnamefont{Geim}},
  \bibinfo{author}{\bibfnamefont{S.~V.} \bibnamefont{Morozov}},
  \bibinfo{author}{\bibfnamefont{E.~W.} \bibnamefont{Hill}},
  \bibinfo{author}{\bibfnamefont{P.}~\bibnamefont{Blake}},
  \bibinfo{author}{\bibfnamefont{M.~I.} \bibnamefont{Katsnelson}},
  \bibnamefont{and} \bibinfo{author}{\bibfnamefont{K.~S.}
  \bibnamefont{Novoselov}}, \bibinfo{journal}{Nat. Mater.}
  \textbf{\bibinfo{volume}{6}}, \bibinfo{pages}{652} (\bibinfo{year}{2007}).

\bibitem[{\citenamefont{Bolotin et~al.}(2008)\citenamefont{Bolotin, Sikes,
  Jiang, Klima, Fudenberg, Hone, Kim, and Stormer}}]{bolotin08}
\bibinfo{author}{\bibfnamefont{K.}~\bibnamefont{Bolotin}},
  \bibinfo{author}{\bibfnamefont{K.}~\bibnamefont{Sikes}},
  \bibinfo{author}{\bibfnamefont{Z.}~\bibnamefont{Jiang}},
  \bibinfo{author}{\bibfnamefont{M.}~\bibnamefont{Klima}},
  \bibinfo{author}{\bibfnamefont{G.}~\bibnamefont{Fudenberg}},
  \bibinfo{author}{\bibfnamefont{J.}~\bibnamefont{Hone}},
  \bibinfo{author}{\bibfnamefont{P.}~\bibnamefont{Kim}}, \bibnamefont{and}
  \bibinfo{author}{\bibfnamefont{H.~L.} \bibnamefont{Stormer}},
  \bibinfo{journal}{Solid State Commun.} \textbf{\bibinfo{volume}{143}},
  \bibinfo{pages}{351} (\bibinfo{year}{2008}).

\bibitem[{\citenamefont{Avouris et~al.}(2007)\citenamefont{Avouris, Chen, and
  Perbeinos}}]{avouris07}
\bibinfo{author}{\bibfnamefont{P.}~\bibnamefont{Avouris}},
  \bibinfo{author}{\bibfnamefont{Z.}~\bibnamefont{Chen}}, \bibnamefont{and}
  \bibinfo{author}{\bibfnamefont{V.}~\bibnamefont{Perbeinos}},
  \bibinfo{journal}{Nature Nanotech} \textbf{\bibinfo{volume}{2}},
  \bibinfo{pages}{605} (\bibinfo{year}{2007}).

\bibitem[{\citenamefont{Novoselov}(2007)}]{novolesov07}
\bibinfo{author}{\bibfnamefont{K.}~\bibnamefont{Novoselov}},
  \bibinfo{journal}{Nat. Mater.} \textbf{\bibinfo{volume}{6}},
  \bibinfo{pages}{720} (\bibinfo{year}{2007}).

\bibitem[{\citenamefont{Meyer et~al.}(2008)\citenamefont{Meyer, Girit, Crommie,
  and Zettl}}]{meyer08}
\bibinfo{author}{\bibfnamefont{J.~C.} \bibnamefont{Meyer}},
  \bibinfo{author}{\bibfnamefont{C.~O.} \bibnamefont{Girit}},
  \bibinfo{author}{\bibfnamefont{M.~F.} \bibnamefont{Crommie}},
  \bibnamefont{and} \bibinfo{author}{\bibfnamefont{A.}~\bibnamefont{Zettl}},
  \bibinfo{journal}{Appl. Phys. Lett.} \textbf{\bibinfo{volume}{92}},
  \bibinfo{pages}{123110} (\bibinfo{year}{2008}).

\bibitem[{\citenamefont{Fischbein and Drndic}(2008)}]{fischbein08}
\bibinfo{author}{\bibfnamefont{M.~D.} \bibnamefont{Fischbein}}
  \bibnamefont{and} \bibinfo{author}{\bibfnamefont{M.}~\bibnamefont{Drndic}},
  \bibinfo{journal}{Appl. Phys. Lett.} \textbf{\bibinfo{volume}{93}},
  \bibinfo{eid}{113107} (\bibinfo{year}{2008}).

\bibitem[{\citenamefont{Son et~al.}(2006)\citenamefont{Son, Cohen, and
  Louie}}]{louie06}
\bibinfo{author}{\bibfnamefont{Y.~W.} \bibnamefont{Son}},
  \bibinfo{author}{\bibfnamefont{M.~L.} \bibnamefont{Cohen}}, \bibnamefont{and}
  \bibinfo{author}{\bibfnamefont{S.~G.} \bibnamefont{Louie}},
  \bibinfo{journal}{Phys. Rev. Lett.} \textbf{\bibinfo{volume}{97}},
  \bibinfo{pages}{216803} (\bibinfo{year}{2006}).

\bibitem[{\citenamefont{Slonczewski and Weiss}(1958)}]{Weiss58}
\bibinfo{author}{\bibfnamefont{J.~C.} \bibnamefont{Slonczewski}}
  \bibnamefont{and} \bibinfo{author}{\bibfnamefont{P.~R.} \bibnamefont{Weiss}},
  \bibinfo{journal}{Phys. Rev.} \textbf{\bibinfo{volume}{109}},
  \bibinfo{pages}{272} (\bibinfo{year}{1958}).

\bibitem[{\citenamefont{Inui et~al.}(1994)\citenamefont{Inui, Trugman, and
  Abrahams}}]{Inui94}
\bibinfo{author}{\bibfnamefont{M.}~\bibnamefont{Inui}},
  \bibinfo{author}{\bibfnamefont{S.~A.} \bibnamefont{Trugman}},
  \bibnamefont{and} \bibinfo{author}{\bibfnamefont{E.}~\bibnamefont{Abrahams}},
  \bibinfo{journal}{Phys. Rev. B} \textbf{\bibinfo{volume}{49}},
  \bibinfo{pages}{3190} (\bibinfo{year}{1994}).

\bibitem[{\citenamefont{Pereira et~al.}(2006)\citenamefont{Pereira, Guinea,
  {Lopes dos Santos}, Peres, and {Castro Neto}}}]{Pereira2006}
\bibinfo{author}{\bibfnamefont{V.~M.} \bibnamefont{Pereira}},
  \bibinfo{author}{\bibfnamefont{F.}~\bibnamefont{Guinea}},
  \bibinfo{author}{\bibfnamefont{J.}~\bibnamefont{{Lopes dos Santos}}},
  \bibinfo{author}{\bibfnamefont{N.}~\bibnamefont{Peres}}, \bibnamefont{and}
  \bibinfo{author}{\bibfnamefont{A.}~\bibnamefont{{Castro Neto}}},
  \bibinfo{journal}{Phys. Rev. Lett.} \textbf{\bibinfo{volume}{96}},
  \bibinfo{pages}{036801} (\bibinfo{year}{2006}).

\bibitem[{\citenamefont{Casolo et~al.}(2009)\citenamefont{Casolo, L{\o}vvik,
  Martinazzo, and Tantardini}}]{casolo09}
\bibinfo{author}{\bibfnamefont{S.}~\bibnamefont{Casolo}},
  \bibinfo{author}{\bibfnamefont{O.~M.} \bibnamefont{L{\o}vvik}},
  \bibinfo{author}{\bibfnamefont{R.}~\bibnamefont{Martinazzo}},
  \bibnamefont{and} \bibinfo{author}{\bibfnamefont{G.~F.}
  \bibnamefont{Tantardini}}, \bibinfo{journal}{J. Chem. Phys.}
  \textbf{\bibinfo{volume}{130}}, \bibinfo{pages}{054704}
  (\bibinfo{year}{2009}).

\bibitem[{\citenamefont{Naumis}(2007)}]{naumis07}
\bibinfo{author}{\bibfnamefont{G.~G.} \bibnamefont{Naumis}},
  \bibinfo{journal}{Phys. Rev. B} \textbf{\bibinfo{volume}{76}},
  \bibinfo{pages}{153403} (\bibinfo{year}{2007}).

\end{thebibliography}

\end{document}